\documentclass[conference]{IEEEtran}
\ifCLASSINFOpdf
\else
\fi
\usepackage[cmex10]{amsmath}
\usepackage[tight,footnotesize]{subfigure}
\usepackage{graphicx}
\usepackage{color}

\begin{document}
\bstctlcite{IEEEexample:BSTcontrol}
%
\title{Extension of Modularity Density for Overlapping Community Structure}

\author{\IEEEauthorblockN{Mingming Chen}
\IEEEauthorblockA{Department of Computer Science\\
Rensselaer Polytechnic Institute\\
110 8th Street, Troy, NY 12180\\
Email: chenm8@rpi.edu}
\and
\IEEEauthorblockN{Konstantin~Kuzmin}
\IEEEauthorblockA{Department of Computer Science\\
Rensselaer Polytechnic Institute\\
110 8th Street, Troy, NY 12180\\
Email: kuzmik@rpi.edu}
\and
\IEEEauthorblockN{Boleslaw K. Szymanski}
\IEEEauthorblockA{Department of Computer Science\\
Rensselaer Polytechnic Institute\\
110 8th Street, Troy, NY 12180\\
Email: szymab@rpi.edu}
}

\maketitle

\begin{abstract}
Modularity is widely used to effectively measure the strength of the disjoint community structure found by community detection algorithms. Although several overlapping extensions of modularity were proposed to measure the quality of overlapping community structure, there is lack of systematic comparison of different extensions. To fill this gap, we overview overlapping extensions of modularity to select the best. In addition, we extend the \textit{Modularity Density} metric to enable its usage for overlapping communities. The experimental results on four real networks using overlapping extensions of modularity, overlapping modularity density, and six other community quality metrics show that the best results are obtained when the product of the belonging coefficients of two nodes is used as the belonging function. Moreover, our experiments indicate that overlapping modularity density is a better measure of the quality of overlapping community structure than other metrics considered.
\end{abstract}

\IEEEpeerreviewmaketitle

\section{Introduction}
Many networks, including Internet, citation networks, transportation networks, e-mail networks, and social and biochemical networks, display community structure which identifies groups of nodes within which connections are denser than between them \cite{UWDModularity}. Detecting and characterizing such community structure, which is known as community detection, is one of the fundamental issues in the study of network systems. Community detection has been shown to reveal latent yet meaningful structure in networks such as groups in online and contact-based social networks, functional modules in protein-protein interaction networks, groups of customers with similar interests in online retailer user networks, groups of scientists in interdisciplinary collaboration networks, etc. \cite{CommunityReport}.

In the last decade, the most popular community detection method, proposed by Newman \cite{NewmanGreedy}, has been to maximize the quality metric known as modularity \cite{UWDModularity,PNASModularity} over all the possible partitions of a network. This metric measures the difference between the fraction of all edges that are within the actual community and such a fraction of edges that would be inside the community in a randomized graph with the same number of nodes and the same degree sequence. It is widely used to measure the strength of community structures discovered by community detection algorithms.

Newman's modularity can only be used to measure the quality of disjoint communities. However, it is more realistic to expect that nodes in real networks belong to more than one community, resulting in overlapping communities \cite{Overlapping_survey}. Therefore, several overlapping extensions of modularity (\cite{fuzzy_cmeans_Qov,fuzzy_opt_Qov,Qov_num_coms,Qov_max_clique,Qov_node_strength,Qov_edge_degree,OverlappingQ_survey}) were proposed to measure the quality of overlapping community structure. Yet, to date no attempts have been made to systematically compare different overlapping extensions and propose metric selection criteria for different types of networks. In this paper, we consider several overlapping extensions of modularity and test their quality on four real networks. We also extend \textit{Modularity Density} \cite{QdsConference,QdsJournal} for overlapping communities following the same principles used by the overlapping extensions of modularity. Finally, we make a comparison of different overlapping extensions of modularity with overlapping modularity density.

We conducted experiments on four real-world networks using overlapping extensions of modularity, overlapping modularity density, and six other metrics (the number of Intra-edges, Intra-density, Contraction, the number of Inter-edges, Expansion, and Conductance). The results show that selecting the product of the belonging coefficients of two nodes as a belonging function for overlapping extensions yields better results on these networks than using other belonging functions. Moreover, the results imply that overlapping modularity density is better than other metrics considered for measuring the quality of overlapping community structures.

\section{Modularity}
\label{sec:related_work}

\subsection{Newman's Modularity}
\label{subsec:modularity}
Newman's modularity \cite{UWDModularity,PNASModularity} for unweighted and undirected networks is defined as the difference between the fractions of the actual and expected (in a randomized graph with the same number of nodes and the same degree sequence) number of edges within the community. A larger value of modularity means a stronger community structure. For the given community partition of a network $G=(V,E)$ with $|E|$ edges, modularity ($Q$) \cite{UWDModularity} is given by:
\begin{equation}
\label{eq:uwdmodularity}
Q=\sum_{c \in C} \left[\frac{|E_{c}^{in}|}{|E|}-\left(\frac{2|E_{c}^{in}|+|E_{c}^{out}|}{2|E|}\right)^2\right],
\end{equation}
where $C$ is the set of all the communities, $c$ is a specific community in $C$, $|E_{c}^{in}|$ is the number of edges between nodes within community $c$, and $|E_{c}^{out}|$ is the number of edges from the nodes in community $c$ to the nodes outside $c$.

Modularity can also be expressed as \cite{PNASModularity}:
\begin{equation}
\label{eq:spectral_Q}
Q=\frac{1}{2|E|} \sum\limits_{ij} \left[A_{ij} - \frac{k_i k_j}{2|E|}\right] \delta_{{c_i}, {c_j}},
\end{equation}
where $k_i$ is the degree of node $i$, $A_{ij}$ is an element of the adjacency matrix between node $i$ and node $j$, $\delta_{{c_i}, {c_j}}$ is the Kronecker delta symbol, and $c_i$ is the label of the community to which node $i$ is assigned.

\subsection{Overlapping Definition of Modularity}
\label{sec:ov_Q}
Newman's modularity is used to measure the quality of disjoint community structure of a network. However, it is more realistic that nodes in networks belong to more than one community, resulting in overlapping communities \cite{Overlapping_survey}. For instance, a researcher may be active in several research areas, and a node in biological networks might have multiple functions. It is also quite common that people in social networks are naturally characterized by multiple community memberships depending on their families, friends, professions, etc. For this reason, discovering overlapping communities became very popular in the last few years. Several overlapping extensions of modularity \cite{fuzzy_cmeans_Qov,fuzzy_opt_Qov,Qov_num_coms,Qov_max_clique,Qov_node_strength,Qov_edge_degree,OverlappingQ_survey} were proposed to measure the quality of overlapping community structure. These extensions are described below.

If communities overlap, each node can belong to multiple communities, although the strength of this connection can generally be different for different communities. Given a set of overlapping communities $C=\{c_1, c_2, ..., c, ..., c_{|C|}\}$ in which a node may belong to more than one of them, a vector of \textit{belonging coefficients} $(a_{i,c_1}, a_{i,c_2}, ..., a_{i,c}, ..., a_{i,c_{|C|}})$ \cite{fuzzy_opt_Qov,Qov_edge_degree} can be assigned to each node $i$ in the network. A belonging coefficient $a_{i,c}$ measures the strength of association between node $i$ and community $c$. Without loss of generality, the following constraints are assumed to hold:
\begin{equation}
0 \le a_{i,c} \le 1~~\forall i \in V, \forall c \in C~~\text{and}~~\sum_{c \in C} a_{i,c} = 1.
\end{equation}

Zhang et al. \cite{fuzzy_cmeans_Qov} proposed an extended modularity which uses the average of the belonging coefficients of two nodes to measure the quality of overlapping community structure:
\begin{equation}
\label{eq:Qov_Z}
Q_{ov}^Z=\sum_{c \in C}\left[ \frac{|E_{c}^{in}|}{|E|} - \left( \frac{2|E_{c}^{in}|+|E_{c}^{out}|}{2|E|} \right)^2 \right],
\end{equation}
where $|E_{c}^{in}|=\frac{1}{2}\sum_{i,j \in c}\frac{a_{i,c}+a_{j,c}}{2}A_{ij}$, $|E_{c}^{out}|=\sum_{i \in c, j \in V-c}\frac{a_{i,c}+(1-a_{j,c})}{2}A_{ij}$, and $|E|=\frac{1}{2}\sum_{ij}A_{ij}$. For the case of disjoint communities, $Q_{ov}^Z$ reduces exactly to Newman's modularity ($Q$) given by Equation~(\ref{eq:uwdmodularity}).

Nepusz et al. \cite{fuzzy_opt_Qov} considered the belonging coefficient $a_{i,c}$ as the probability of the event that node $i$ is in community $c$. Then, the probability of the event that node $i$ belongs to the same communities as node $j$ is the dot product of their membership vectors denoted as $s_{ij}=\sum_{c \in C} a_{i,c}a_{j,c}$.
The authors also adopted $s_{ij}$ as the similarity measure between nodes $i$ and $j$. By replacing $\delta_{c_i,c_j}$ in Equation~(\ref{eq:spectral_Q}) with the similarity measure $s_{ij}$ defined above, they proposed a fuzzified variant of modularity:
\begin{equation}
\label{eq:Qov_F}
\begin{split}
Q_{ov}^{F}&=\frac{1}{2|E|} \sum\limits_{ij} \left[A_{ij} - \frac{k_i k_j}{2|E|}\right] s_{ij} \\
&=\frac{1}{2|E|} \sum_{c \in C} \sum\limits_{i,j \in c} \left[A_{ij} - \frac{k_i k_j}{2|E|}\right]a_{i,c}a_{j,c}.
\end{split}
\end{equation}
In case communities are disjoint, there exists only one community $c$ for every node $i$ with $a_{i,c}=1$. Then, the fuzzified modularity ($Q_{ov}^{F}$) reduces to exactly the original modularity ($Q$) described in Equation~(\ref{eq:spectral_Q}).

Shen et al. \cite{Qov_num_coms} proposed an extension of modularity for overlapping community structure with the same definition to Equation~(\ref{eq:Qov_F}) but defined the belonging coefficients of node $i$ to be the reciprocal of the number of communities to which it belongs:
\begin{equation}
\label{eq:num_com_factor}
a_{i,c}=\frac{1}{O_i},
\end{equation}
where $O_i$ is the number of communities containing node $i$. Then, the extended modularity for overlapping community structure is given by:
\begin{equation}
\label{eq:Qov_E}
\begin{split}
Q_{ov}^{E}&=\frac{1}{2|E|} \sum_{c \in C} \sum\limits_{i,j \in c} \left[A_{ij} - \frac{k_i k_j}{2|E|}\right]a_{i,c}a_{j,c} \\
&=\frac{1}{2|E|} \sum_{c \in C} \sum\limits_{i,j \in c} \left[A_{ij} - \frac{k_i k_j}{2|E|}\right]\frac{1}{O_iO_j}.
\end{split}
\end{equation}
For disjoint community structure, $Q_{ov}^E$ reduces to the original modularity ($Q$) described in Equation~(\ref{eq:spectral_Q}).

Shen et al. \cite{Qov_max_clique} proposed another extension of modularity for overlapping communities with the same definition to Equation~(\ref{eq:Qov_F}). In this case, the belonging coefficient of node $i$ to community $c$ is defined as:
\begin{equation}
\label{eq:max_clique_factor}
a_{i,c}=\frac{1}{a_i}\sum_{k\in c} \frac{M_{ik}^c}{M_{ik}} A_{ik},
\end{equation}
where $M_{ik}$ denotes the number of maximal cliques in the network containing edge $(i,k)$, $M_{ik}^c$ is the number of maximal cliques in community $c$ that contains edge $(i,k)$, and $a_{i}=\sum_{c \in C} \sum_{k \in c} \frac{M_{ik}^c}{M_{ik}} A_{ik}$ is a normalization term.
The maximal clique is a clique that is not a subset of any other cliques. Then, the extended modularity for overlapping community structure is given by:
\begin{equation}
\label{eq:Qov_C}
Q_{ov}^{C}=\frac{1}{2|E|} \sum_{c \in C} \sum\limits_{i,j \in c} \left[A_{ij} - \frac{k_i k_j}{2|E|}\right]a_{i,c}a_{j,c}.
\end{equation}
Note that for disjoint communities, this new extension also reduces to Newman's modularity shown in Equation~(\ref{eq:spectral_Q}).

Chen et al. \cite{Qov_node_strength} also proposed another extension of modularity with the same definition to Equation~(\ref{eq:Qov_F}) but with the belonging coefficient defined as:
\begin{equation}
\label{eq:node_strength_factor}
a_{i,c}=\frac{\sum_{k \in c} A_{ik}}{\sum_{c' \in C_i} \sum_{k \in c'} A_{ik}},
\end{equation}
where $C_i$ is the set of communities to which node $i$ belongs. It measures how tightly node $i$ connects to community $c$. Consequently, the extended definition of modularity for overlapping community structure is given by:
\begin{equation}
\label{eq:Qov_O}
\begin{split}
&Q_{ov}^{O}=\frac{1}{2|E|} \sum_{c \in C} \sum\limits_{i,j \in c} \left[A_{ij} - \frac{k_i k_j}{2|E|}\right]a_{i,c}a_{j,c} \\
&=\frac{1}{2|E|} \sum_{c \in C} \sum\limits_{i,j \in c} \left[A_{ij} - \frac{k_i k_j}{2|E|}\right]\frac{\sum\limits_{k \in c} A_{ik}}{\sum\limits_{c' \in C_i} \sum\limits_{k \in c'} A_{ik}} \frac{\sum\limits_{k \in c} A_{jk}}{\sum\limits_{c' \in C_j} \sum\limits_{k \in c'} A_{jk}}.
\end{split}
\end{equation}
Still, for disjoint community structure, $Q_{ov}^{O}$ reduces to the original modularity given by Equation~(\ref{eq:spectral_Q}).

Unlike the node-based extensions of modularity presented above, Nicosia et al. \cite{Qov_edge_degree} proposed an edge-based extension of modularity for overlapping communities. In this case, the belonging coefficients represent how edges are assigned to communities. The belonging coefficient for edge $l=(i,j)$ to community $c$ is $\beta_{l(i,j),c}=F(a_{i,c}, a_{j,c})$,
where $F(a_{i,c}, a_{j,c})$ could be any function (product, average, or maximum) of $a_{i,c}$ and $a_{j,c}$. After trying several different functions, the authors stated that the best $F$ is a two-dimensional logistic function:
\begin{equation}
\label{eq:Qov_L_bf}
F(a_{i,c}, a_{j,c}) = \frac{1}{(1+e^{-f(a_{i,c})})(1+e^{-f(a_{j,c})})},
\end{equation}
where $f(a_{i,c})$ is a simple linear scaling function $f(x)=2px-p, p \in R$.
In papers \cite{Overlapping_survey,ov_lpa}, $p$ was selected to be $30$. Then, the expected belonging coefficient of any edge $l=(i,k)$ starting from node $i$ in community $c$ is given by $\beta_{l(i,k),c}^e=\frac{1}{|V|}\sum_{k \in V}\beta_{l(i,k),c}$ running over all nodes in the network. Accordingly, the expected belonging coefficient of any edge $l=(k,j)$ pointing to node $j$ in community $c$ is defined as $\beta_{l(k,j),c}^e=\frac{1}{|V|}\sum_{k \in V}\beta_{l(k,j),c}$. Then, the edge-based extension of modularity is given by:
\begin{equation}
\label{eq:Qov_edge}
\begin{split}
Q_{ov}^L&=\frac{1}{2|E|} \sum_{c \in C} \sum\limits_{i,j \in c} \left[r_{ijc}A_{ij} - s_{ijc}\frac{k_i k_j}{2|E|}\right] \\
&=\frac{1}{2|E|} \sum_{c \in C} \sum\limits_{i,j \in c} \left[\beta_{l(i,j),c}A_{ij} - \frac{\beta_{l(i,k),c}^e k_i \beta_{l(k,j),c}^e k_j}{2|E|}\right],
\end{split}
\end{equation}
where
\begin{equation}
r_{ijc}=\beta_{l(i,j),c}=F(a_{i,c}, a_{j,c})
\end{equation}
and
\begin{equation}
\begin{split}
s_{ijc}=\frac{\sum_{k \in V}F(a_{i,c}, a_{k,c})\sum_{k \in V}F(a_{k,c}, a_{j,c})}{|V|^2}.
\end{split}
\end{equation}
In $Q_{ov}^L$, $r_{ijc}$ is used as the weight corresponding to the probability of the observed edge $l=(i,j)$, while $s_{ijc}$ is used as the weight of the probability of an edge from node $i$ to node $j$ in the null model. Note that although for disjoint communities $F(a_{i,c}, a_{j,c})$ is practically 1 when both $a_{i,c}$ and $a_{j,c}$ are equal to 1, $Q_{ov}^L$ does not exactly reduce to the original modularity given by Equation~(\ref{eq:spectral_Q}).

Generally, there are two categories of overlapping community structures: crisp (non-fuzzy) and fuzzy \cite{OverlappingQ_survey}. For crisp overlapping community structure, each node belongs to one or more communities but without the corresponding belonging coefficients. That is, the relationship between a node and a community is binary: a node either belongs to a community or it does not. For fuzzy overlapping community structure, each node can be a member of multiple communities, but in general the values of belonging coefficients are different. Fuzzy overlapping can be easily transformed to crisp overlapping with a threshold parameter. Namely, if the belonging coefficient of node $i$ to community $c$ is larger than the value of the threshold, then node $i$ stays in community $c$. Otherwise, node $i$ is deleted from community $c$. Crisp overlapping can be converted to fuzzy overlapping by calculating the value of the belonging coefficient using Equations~(\ref{eq:num_com_factor}), (\ref{eq:max_clique_factor}), or (\ref{eq:node_strength_factor}). However, calculating the belonging coefficient using Equation~(\ref{eq:max_clique_factor}) is computationally expensive since it is first necessary to find all the maximal cliques of the network. Hence, in this paper we only consider Equation~(\ref{eq:num_com_factor}) and Equation~(\ref{eq:node_strength_factor}) when converting crisp overlapping to fuzzy overlapping.

Now, we give two general definitions, $Q_{ov}$ and $Q_{ov}'$, for node-based extensions of modularity. First, $Q_{ov}$ is given by:
\begin{equation}
\label{eq:Qov}
Q_{ov}=\sum_{c \in C}\left[ \frac{|E_{c}^{in}|}{|E|} - \left( \frac{2|E_{c}^{in}|+|E_{c}^{out}|}{2|E|} \right)^2 \right],
\end{equation}
where $|E_{c}^{in}|=\frac{1}{2}\sum_{i,j \in c} f(a_{i,c}, a_{j,c})A_{ij}$, $|E_{c}^{out}|=\sum_{i \in c} \sum_{\substack{c' \in C \\ c' \ne c \\ j \in c' }}f(a_{i,c}, a_{j,c'})A_{ij}$, and $|E|=\frac{1}{2}\sum_{ij}A_{ij}$. The belonging function $f(a_{i,c}, a_{j,c})$ can be the average or product of $a_{i,c}$ and $a_{j,c}$. That is, $f(a_{i,c}, a_{j,c})=\frac{a_{i,c}+a_{j,c}}{2}$ or $f(a_{i,c}, a_{j,c})=a_{i,c}a_{j,c}$. Clearly, $Q_{ov}$ with $f(a_{i,c}, a_{j,c})=\frac{a_{i,c}+a_{j,c}}{2}$ is very similar to $Q_{ov}^Z$ in Equation~(\ref{eq:Qov_Z}). Second, $Q_{ov}'$ is given by:
\begin{equation}
Q_{ov}'=\frac{1}{2|E|} \sum_{c \in C} \sum\limits_{i,j \in c} \left[A_{ij} - \frac{k_i k_j}{2|E|}\right] f(a_{i,c}, a_{j,c}).
\end{equation}
where $f(a_{i,c}, a_{j,c})$ is the same as that in Equation~(\ref{eq:Qov}). It is worth noting that $Q_{ov}'$ with the belonging function $f(a_{i,c}, a_{j,c})=a_{i,c}a_{j,c}$ is actually the same as $Q_{ov}^{F}$ in Equation~(\ref{eq:Qov_F}), $Q_{ov}^{E}$ in Equation~(\ref{eq:Qov_E}), $Q_{ov}^{C}$ in Equation~(\ref{eq:Qov_C}), and $Q_{ov}^{O}$ in Equation~(\ref{eq:Qov_O}). The only difference between these formulas is how the value of $a_{i,c}$ is calculated.

It is easy to prove that $Q_{ov}$ is equivalent to $Q_{ov}'$ when $f(a_{i,c}, a_{j,c})=a_{i,c}a_{j,c}$. From the definition of $Q_{ov}$, we know that
$|E_{c}^{in}|=\frac{1}{2}\sum_{i,j \in c} a_{i,c}a_{j,c}A_{ij}$ which is in fact the same as the first term of $Q_{ov}'$. Moreover, it is easy to show that $\left( 2|E_{c}^{in}|+|E_{c}^{out}|\right)^2=\sum_{i,j \in c} k_ik_ja_{i,c}a_{j,c}$.
Hence, the second term of $Q_{ov}$ is the same as the second term of $Q_{ov}'$. Similarly, it can be shown that $Q_{ov}$ is not equal to $Q_{ov}'$ when $f(a_{i,c}, a_{j,c})=\frac{a_{i,c}+a_{j,c}}{2}$.

\section{Modularity Density}
\subsection{Modularity Density for Disjoint Communities}
Chen et al. \cite{QdsConference,QdsJournal} proposed \textit{Modularity Density} which simultaneously addresses two opposite yet coexisting problems of Newman's modularity: in some cases, it tends to favor small communities over large ones while in others, large communities over small ones. The latter tendency is known in the literature as the resolution limit problem of modularity \cite{ResolutionLimit}. Modularity density mixes two additional components, \textit{Split Penalty} and the community density, into Newman's modularity given in Equation~(\ref{eq:uwdmodularity}). Split penalty is the fraction of edges that connect nodes of different communities. Community density includes internal community density and pair-wise community density. The definition of \textit{Modularity Density} ($Q_{ds}$) for unweighted and undirected networks is given by:
\begin{equation}
\label{eq:Qds}
\begin{split}
&Q_{ds}=\sum_{c \in C} \biggl[\frac{|E_{c}^{in}|}{|E|}d_{c}-\left(\frac{2|E_{c}^{in}|+|E_{c}^{out}|}{2|E|}d_{c}\right)^2 \\
& ~~~~~~~~~~~~~-\sum_{\substack{c' \in C \\ c' \ne c}}\frac{|E_{c,c'}|}{2|E|}d_{c,c'}\biggr], \\
\end{split}
\end{equation}
where $d_{c}=\frac{2|E_{c}^{in}|}{|c|(|c|-1)}$ is the internal density of community $c$, and $d_{c,c'}=\frac{|E_{c,c'}|}{|c||c'|}$ is the pair-wise density between community $c$ and community $c'$.

\subsection{Modularity Density for Overlapping Communities}
According to $Q_{ov}$ in Equation~(\ref{eq:Qov}), we extend $Q_{ds}$ for overlapping community structure as:
\begin{equation}
\label{eq:Qov_ds}
\begin{split}
&Q_{ds}^{ov}=\sum_{c \in C} \biggl[\frac{|E_{c}^{in}|}{|E|}d_{c}-\left(\frac{2|E_{c}^{in}|+|E_{c}^{out}|}{2|E|}d_{c}\right)^2 \\
& ~~~~~~~~~~~~~-\sum_{\substack{c' \in C \\ c' \ne c}}\frac{|E_{c,c'}|}{2|E|}d_{c,c'}\biggr], \\
&d_c = \frac{2|E_{c}^{in}|}{\sum_{i,j \in c, i \ne j} f(a_{i,c}, a_{j,c})}, \\
&d_{c,c'}=\frac{|E_{c,c'}|}{\sum_{i \in c, j \in c'} f(a_{i,c}, a_{j,c'})},
\end{split}
\end{equation}
where $|E_{c}^{in}|=\frac{1}{2}\sum_{i,j \in c} f(a_{i,c}, a_{j,c})A_{ij}$, $|E_{c}^{out}|=\sum\limits_{i \in c} \sum\limits_{\substack{c' \in C \\ c' \ne c \\ j \in c' }}f(a_{i,c}, a_{j,c'})A_{ij}$, $|E_{c,c'}|=\sum_{i \in c, j \in c'} f(a_{i,c}, a_{j,c'})A_{ij}$, and $|E|=\frac{1}{2}\sum\limits_{ij}A_{ij}$. Belonging function $f(a_{i,c}, a_{j,c})$ can be the product or average of $a_{i,c}$ and $a_{j,c}$. For disjoint communities, $Q_{ds}^{ov}$ reduces exactly to $Q_{ds}$ given by Equation~(\ref{eq:Qds}). Notice that we do not extend modularity density based on $Q_{ov}^L$ since it is too complicated and far from intuitive.

\begin{table*}[!t]
\caption{The values of the metrics with the first version of belonging coefficient and the first version of belonging function on Zachary's karate club network.}
\label{tab:karate_1_1}
\vspace{-0.9em}
\centering
\begin{tabular}{c||c|c|c|c|c|c|c|c|c|c|c}
\hline \hline
     SLPA threshold $r$ & 0.01 & 0.05 & 0.1 & 0.15 & 0.2 & 0.25 & 0.3 & 0.35 & 0.4 & 0.45 & 0.5 \\
\hline
     $Q_{ov}$ & 0.1247 & 0.1791	& 0.2386 & 0.269 & 0.2815 & 0.3199 & 0.3368 & 0.3542 & 0.3676 & 0.3751 & \textcolor{red}{\textbf{\emph{0.3785}}} \\
\hline
     $Q_{ov}^L$ & 0.6121 & 0.6404 & 0.6359 & 0.6474 & 0.6859 & 0.7024 & 0.7054 & 0.6983 & 0.7088 & \textcolor{red}{\textbf{\emph{0.7169}}} & 0.7165\\
\hline
     $Q_{ds}^{ov}$ & 0.1212 & 0.1389 & 0.1528 & 0.1606 & 0.1687 & 0.1822 & 0.185 & 0.1814 & 0.1878 & 0.1939 & \textcolor{red}{\textbf{\emph{0.1946}}} \\
\hline
    \# \textit{Intra-edges} & \textcolor{red}{\textbf{\emph{28.6056}}} & 27.1188 & 27.3083 & 27.2154 & 25.9083 & 24.5333 & 24.2167 & 22.2196 & 24.4779 & 24.7396 & 24.6083 \\
\hline
     \textit{Intra-density} & 0.2615 & 0.2859 & 0.305 & 0.3106 & 0.3101 & 0.3541 & 0.3603 & \textcolor{red}{\textbf{\emph{0.4013}}} & 0.3596 & 0.3491 & 0.35 \\
\hline
     \textit{Contraction} & \textcolor{red}{\textbf{\emph{3.9861}}} & 3.9026 & 3.7906 & 3.7864 & 3.7448 & 3.6468 & 3.5955 & 3.4554 & 3.5413 & 3.5562 & 3.5374 \\
\hline
    \# \textit{Inter-edges} & 25.8806 & 22.8458 & 17.9383 & 16.2892 & 16.7083 & 14.3083 & 13.1583 & 12.3008 & 11.3375 & 10.8208 & \textcolor{red}{\textbf{\emph{10.5833}}} \\
\hline
     \textit{Expansion} & 2.2719 & 2.04559 & 1.7042 & 1.5569 & 1.4758 & 1.3831 & 1.2819 & 1.374 & 1.1055 & 0.9607 & \textcolor{red}{\textbf{\emph{0.9404}}} \\
\hline
     \textit{Conductance} & 0.333 & 0.3206 & 0.2842 & 0.2682 & 0.2677 & 0.2628 & 0.2523 & 0.2704 & 0.23 & 0.2141 & \textcolor{red}{\textbf{\emph{0.2121}}} \\
\hline \hline
\end{tabular}
\vspace{-0.2em}
\end{table*}

\begin{table*}[!t]
\caption{The values of the metrics with the first version of belonging coefficient and the second version of belonging function on Zachary's karate club network.}
\label{tab:karate_1_2}
\vspace{-0.9em}
\centering
\begin{tabular}{c||c|c|c|c|c|c|c|c|c|c|c}
\hline \hline
     SLPA threshold $r$ & 0.01 & 0.05 & 0.1 & 0.15 & 0.2 & 0.25 & 0.3 & 0.35 & 0.4 & 0.45 & 0.5 \\
\hline
     $Q_{ov}$ & 0.2873 & 0.3073 & 0.3159 & 0.3277 & 0.348 & 0.3635 & 0.3665 & 0.3693 & 0.3744 & \textcolor{red}{\textbf{\emph{0.3787}}} & 0.3785 \\
\hline
     $Q_{ov}^L$ & 0.6121 & 0.6404 & 0.6359 & 0.6474 & 0.6859 & 0.7024 & 0.70537 & 0.6983 & 0.7088 & \textcolor{red}{\textbf{\emph{0.7169}}} & 0.7165 \\
\hline
     $Q_{ds}^{ov}$ & 0.1338 & 0.1467 & 0.1593 & 0.166 & 0.1726 & 0.1844 & 0.1873 & 0.1863 & 0.1898 & \textcolor{red}{\textbf{\emph{0.1948}}} & 0.1946\\
\hline
    \# \textit{Intra-edges} & 24.1854 & 23.3833 & 24.8104 & \textcolor{red}{\textbf{\emph{25.1246}}} & 23.8146 & 23.05 & 23.15 & 21.5625 & 24.1892 & 24.6208 & 24.6083\\
\hline
     \textit{Intra-density} & 0.2749 & 0.2989 & 0.3125 & 0.3182 & 0.3175 & 0.3612 & 0.3675 & \textcolor{red}{\textbf{\emph{0.4053}}} & 0.3612 & 0.3512 & 0.35 \\
\hline
     \textit{Contraction} & 3.2357 & 3.2562 & 3.3278 & 3.3863 & 3.3892 & 3.3634 & 3.381 & 3.2936 & 3.4818 & \textcolor{red}{\textbf{\emph{3.538}}} & 3.5374 \\
\hline
    \# \textit{Inter-edges} & 15.3292 & 14.3333 & 11.9992 & 11.3708 & 12.1708 & 11.1 & 10.9 & 10.695 & 10.6417 & \textcolor{red}{\textbf{\emph{10.5583}}} & 10.5833 \\
\hline
     \textit{Expansion} & 1.3043 & 1.2649 & 1.1108 & 1.0537 & 1.0648 & 1.0482 & 1.0321 & 1.1457 & 1.0199 & \textcolor{red}{\textbf{\emph{0.9388}}} & 0.9404  \\
\hline
     \textit{Conductance} & 0.2884 & 0.2801 & 0.2563 & 0.2426 & 0.2447 & 0.2447 & 0.2404 & 0.2605 & 0.225 & \textcolor{red}{\textbf{\emph{0.2118}}} & 0.2121 \\
\hline \hline
\end{tabular}
\vspace{-0.2em}
\end{table*}

\begin{table*}[!t]
\caption{The values of the metrics with the second version of belonging coefficient and the first version of belonging function on Zachary's karate club network.}
\label{tab:karate_2_1}
\vspace{-0.9em}
\centering
\begin{tabular}{c||c|c|c|c|c|c|c|c|c|c|c}
\hline \hline
     SLPA threshold $r$ & 0.01 & 0.05 & 0.1 & 0.15 & 0.2 & 0.25 & 0.3 & 0.35 & 0.4 & 0.45 & 0.5 \\
\hline
     $Q_{ov}$ & 0.1299 & 0.1822 & 0.2416 & 0.2699 & 0.2832 & 0.3213 & 0.3381 & 0.3551 & 0.3682 & 0.3755 & \textcolor{red}{\textbf{\emph{0.3785}}} \\
\hline
     $Q_{ov}^L$ & 0.6727 & 0.6871 & 0.6696 & 0.6695 & 0.7115 & 0.7158 & 0.7169 & 0.7092 & 0.7142 & \textcolor{red}{\textbf{\emph{0.7199}}} & 0.7165 \\
\hline
     $Q_{ds}^{ov}$ & 0.126 & 0.1423 & 0.1552 & 0.162 & 0.1704 & 0.1833 & 0.1858 & 0.1831 & 0.1887 & 0.1941 & \textcolor{red}{\textbf{\emph{0.1946}}} \\
\hline
    \# \textit{Intra-edges} & \textcolor{red}{\textbf{\emph{28.8916}}} & 27.3525 & 27.4537 & 27.2934 & 26.0103 & 24.612 & 24.2808 & 22.2586 & 24.4972 & 24.7512 & 24.6083 \\
\hline
     \textit{Intra-density} & 0.2639 & 0.2879 & 0.3073 & 0.3123 & 0.3121 & 0.3566 & 0.3622 & \textcolor{red}{\textbf{\emph{0.4017}}} & 0.3601 & 0.3494 & 0.35 \\
\hline
     \textit{Contraction} & \textcolor{red}{\textbf{\emph{4.0207}}} & 3.9292 & 3.8124 & 3.7979 & 3.7612 & 3.6635 & 3.6081 & 3.4598 & 3.5446 & 3.5581 & 3.5374 \\
\hline
    \# \textit{Inter-edges} & 25.3084 & 22.3783 & 17.6476 & 16.1332 & 16.5044 & 14.151 & 13.0301 & 12.2229 & 11.2989 & 10.7976 & \textcolor{red}{\textbf{\emph{10.5833}}} \\
\hline
     \textit{Expansion} & 2.2486 & 2.0434 & 1.7101 & 1.5631 & 1.4967 & 1.4136 & 1.3092 & 1.3879 & 1.108 & 0.9601 & \textcolor{red}{\textbf{\emph{0.9404}}} \\
\hline
     \textit{Conductance} & 0.3274 & 0.3168 & 0.2817 & 0.2671 & 0.2666 & 0.2625 & 0.2523 & 0.2706 & 0.2296 & 0.2138 & \textcolor{red}{\textbf{\emph{0.2121}}} \\
\hline \hline
\end{tabular}
\vspace{-0.2em}
\end{table*}

\begin{table*}[!t]
\caption{The values of the metrics with the second version of belonging coefficient and the second version of belonging function on Zachary's karate club network.}
\label{tab:karate_2_2}
\vspace{-0.9em}
\centering
\begin{tabular}{c||c|c|c|c|c|c|c|c|c|c|c}
\hline \hline
     SLPA threshold $r$ & 0.01 & 0.05 & 0.1 & 0.15 & 0.2 & 0.25 & 0.3 & 0.35 & 0.4 & 0.45 & 0.5 \\
\hline
     $Q_{ov}$ & 0.3013 & 0.3173 & 0.3231 & 0.3311 & 0.353 & 0.3667 & 0.3692 & 0.3716 & 0.3757 & \textcolor{red}{\textbf{\emph{0.3795}}} & 0.3785 \\
\hline
     $Q_{ov}^L$ & 0.6727 & 0.6871 & 0.6696 & 0.6695 & 0.7115 & 0.7158 & 0.7169 & 0.7092 & 0.7142 & \textcolor{red}{\textbf{\emph{0.7199}}} & 0.7165 \\
\hline
     $Q_{ds}^{ov}$ & 0.1435 & 0.154 & 0.1641 & 0.1689 & 0.1761 & 0.1864 & 0.1887 & 0.189 & 0.1914 & \textcolor{red}{\textbf{\emph{0.1953}}} & 0.1946 \\
\hline
    \# \textit{Intra-edges} & 24.7296 & 23.817 & 25.0706 & \textcolor{red}{\textbf{\emph{25.2725}}} & 24.0168 & 23.1996 & 23.2765 & 21.6375 & 24.2284 & 24.6452 & 24.6083 \\
\hline
     \textit{Intra-density} & 0.2829 & 0.3054 & 0.3191 & 0.3237 & 0.3237 & 0.3676 & 0.3717 & \textcolor{red}{\textbf{\emph{0.4061}}} & 0.3624 & 0.352 & 0.35 \\
\hline
     \textit{Contraction} & 3.3015 & 3.3013 & 3.3563 & 3.401 & 3.4075 & 3.369 & 3.3818 & 3.2892 & 3.4835 & \textcolor{red}{\textbf{\emph{3.5411}}} & 3.5374 \\
\hline
    \# \textit{Inter-edges} & 14.2408 & 13.466 & 11.4788 & 11.075 & 11.7663 & 10.8009 & 10.6469 & 10.545 & 10.5631 & \textcolor{red}{\textbf{\emph{10.5095}}} & 10.5833 \\
\hline
     \textit{Expansion} & 1.2211 & 1.2051 & 1.0751 & 1.0338 & 1.0458 & 1.0409 & 1.0271 & 1.1302 & 1.0073 & \textcolor{red}{\textbf{\emph{0.9356}}} & 0.9404 \\
\hline
     \textit{Conductance} & 0.2721 & 0.2687 & 0.2492 & 0.2388 & 0.2401 & 0.2436 & 0.2403 & 0.261 & 0.2242 & \textcolor{red}{\textbf{\emph{0.2111}}} & 0.2121 \\
\hline \hline
\end{tabular}
\vspace{-0.5em}
\end{table*}

\section{Evaluation and Analysis}
\label{sec:evaluation}

From Subsection~\ref{sec:ov_Q}, we know that all node-based overlapping extensions of modularity can be expressed with $Q_{ov}$ in Equation~(\ref{eq:Qov}) using the belonging function $f(a_{i,c}, a_{j,c})=\frac{a_{i,c}+a_{j,c}}{2}$ or $f(a_{i,c}, a_{j,c})=a_{i,c}a_{j,c}$. For the edge-based overlapping extension of modularity ($Q_{ov}^L$), the belonging function is given by Equation~(\ref{eq:Qov_L_bf}). For the overlapping extension of modularity density $(Q_{ds}^{ov})$, the belonging function $f(a_{i,c}, a_{j,c})$ can also be the average or the product of $a_{i,c}$ and $a_{j,c}$. Thus, there are two versions of the belonging function for $Q_{ov}$ and $Q_{ds}^{ov}$. Therefore, we have $Q_{ov}(average)$ with $f(a_{i,c}, a_{j,c})=\frac{a_{i,c}+a_{j,c}}{2}$, $Q_{ov}(product)$ with $f(a_{i,c}, a_{j,c})=a_{i,c}a_{j,c}$, $Q_{ov}^L$ in Equation~(\ref{eq:Qov_edge}), $Q_{ds}^{ov}(average)$ with $f(a_{i,c}, a_{j,c})=\frac{a_{i,c}+a_{j,c}}{2}$, and $Q_{ds}^{ov}(product)$ with $f(a_{i,c}, a_{j,c})=a_{i,c}a_{j,c}$. For fuzzy overlapping community structures, $a_{i,c}$ is given for each node $i$ to community $c$. For crisp overlapping community structures, we can adopt Equation~(\ref{eq:num_com_factor}) and Equation~(\ref{eq:node_strength_factor}) to calculate $a_{i,c}$. Consequently, two versions of the belonging coefficient can be used to convert crisp overlapping to fuzzy overlapping.

We also consider six other community quality metrics: the number of \textit{Intra-edges}, \textit{Intra-density}, \textit{Contraction}, the number of \textit{Inter-edges}, \textit{Expansion}, and \textit{Conductance} \cite{QdsConference,QdsJournal}. These metrics describe how the connectivity structure of a given set of nodes resembles a community. All of them rely on the intuition that communities are sets of nodes with many edges inside them and few edges outside of them. We also extend these metrics to be applicable to overlapping communities. Two versions of the belonging coefficient and two versions of the belonging function are considered for each metric. For fuzzy overlapping community structure, we define the size of a community $c$ as $|c|=\sum_{i \in c}a_{i,c}$.\\
\textbf{The number of \textit{Intra-edges}:} $|E_c^{in}|$; it is the total number of edges in $c$. A large value of this metric is better than a small value in terms of the community quality. \\
\textbf{\textit{Intra-density}:} $d_c$ in Equation~(\ref{eq:Qov_ds}). The larger the value of this metric, the higher the quality of the communities. \\
\textbf{\textit{Contraction}:} $2|E_c^{in}|/|c|$; it measures the average number of edges per node inside the community $c$. A larger value of contraction means better community quality. \\
\textbf{The number of \textit{Inter-edges}:} $|E_c^{out}|$; it is the total number of edges on the boundary of $c$. A small value of this metric is better than a large value in terms of the community quality.  \\
\textbf{\textit{Expansion}:} $|E_c^{out}|/|c|$; it measures the average number of edges (per node) that point outside the community $c$. A smaller value of expansion corresponds to better community structure. \\
\textbf{\textit{Conductance}:} $\frac{|E_c^{out}|}{2|E_c^{in}|+|E_c^{out}|}$; it measures the fraction of the total number of edges that point outside the community. A smaller value of conductance means better community quality.

In this section, we compare different choices of the belonging coefficient and the belonging function to be used for $Q_{ov}$, $Q_{ov}^L$, and $Q_{ds}^{ov}$. Then, we try to determine which of the three overlapping extensions of modularity (two kinds of node-based extensions of modularity and the edge-based extension of modularity) is the best. Finally, we consider the best version of $Q_{ov}$, $Q_{ov}^L$, and $Q_{ds}^{ov}$ and determine which one yields the best results for the networks that we are using. The experiments are done with a community detection algorithm called Speaker-listener Label Propagation Algorithm (SLPA) \cite{SLPA2012} on four real networks which are introduced in the next subsection.

\subsection{Real Networks}
\textbf{Zachary's karate club network} \cite{karate}. It represents the friendships between $34$ members of a karate club at a US university during 2 years. It has 34 nodes and 78 edges.

\textbf{American college football network} \cite{football}. It represents the schedule of games between college football teams in a single season. It has 115 nodes and 613 edges.

\textbf{Jazz musicians network} \cite{Jazz}. It is a network (198 nodes and 2742 edges) of collaborations between jazz musicians.

\textbf{PGP network} \cite{PGPNetwork}. It is the giant component of the network of users of the Pretty-Good-Privacy (PGP) algorithm for secure information interchange. It has 10680 nodes and 24316 edges in total.

\subsection{Experimental Results}
\label{subsec:experiment}
In this subsection, we present the results of performing community detection on four real-world networks described above by using SLPA \cite{SLPA2012} with threshold parameter $r$ varying from 0.01 to 0.5. SLPA gets crisp overlapping communities when $r<0.5$ and gets disjoint communities when $r=0.5$. For each value of threshold $r$, we adopt 10 running samples since the community detection result of SLPA is not deterministic. Then, for the community detection results of SLPA with different values of threshold $r$ on each of the four networks, we calculate the values of $Q_{ov}$, $Q_{ov}^L$, $Q_{ds}^{ov}$, and six community quality metrics with two versions of the belonging coefficient and two versions of the belonging function. For a certain value of $r$, the values of all the metrics are calculated as the average of the  10 samples. For convenience, we denote Equation~(\ref{eq:num_com_factor}) and Equation~(\ref{eq:node_strength_factor}) as the first and the second version of the belonging coefficient, respectively. We also denote the belonging function $f(a_{i,c}, a_{j,c})=\frac{a_{i,c}+a_{j,c}}{2}$ as the first version of the belonging function and $f(a_{i,c}, a_{j,c})=a_{i,c}a_{j,c}$ as the second version of the belonging function. We determine which version of the belonging coefficient and which version of the belonging function are better based on the largest number of quality metrics consistent with each other on determining the best value of threshold $r$ for SLPA on all the networks.

\begin{table*}[!t]
\caption{The values of the metrics with the first version of belonging coefficient and the first version of belonging function on American college football network.}
\label{tab:football_1_1}
\vspace{-0.9em}
\centering
\begin{tabular}{c||c|c|c|c|c|c|c|c|c|c|c}
\hline \hline
     SLPA threshold $r$ & 0.01 & 0.05 & 0.1 & 0.15 & 0.2 & 0.25 & 0.3 & 0.35 & 0.4 & 0.45 & 0.5 \\
\hline
     $Q_{ov}$ & 0.53725 & 0.56603 & 0.57319 & 0.57449 & 0.57517 & 0.57579 & 0.57605 & 0.57634 & \textcolor{red}{\textbf{\emph{0.57641}}} & 0.57636 & 0.57636 \\
\hline
     $Q_{ov}^L$ & 0.66802 & 0.7019 & 0.71254 & 0.71519 & 0.71691 & 0.71811 & 0.71915 & 0.71933 & 0.71941 & \textcolor{red}{\textbf{\emph{0.71943}}} & \textcolor{red}{\textbf{\emph{0.71943}}} \\
\hline
     $Q_{ds}^{ov}$ & 0.3009 & 0.3471 & 0.3618 & 0.366 & 0.369 & 0.3709 & 0.3723 & 0.3735 & 0.3736 & \textcolor{red}{\textbf{\emph{0.3747}}} & \textcolor{red}{\textbf{\emph{0.3747}}} \\
\hline
    \# \textit{Intra-edges} & \textcolor{red}{\textbf{\emph{55.8985}}} & 55.0975 & 54.9343 & 54.8663 & 54.7258 & 54.6529 & 54.4813 & 54.3944 & 54.3606 & 54.245 & 54.245 \\
\hline
     \textit{Intra-density} & 0.5908 & 0.6611 & 0.6805 & 0.6879 & 0.6941 & 0.6968 & 0.7002 & 0.7031 & 0.7031 & \textcolor{red}{\textbf{\emph{0.7061}}} & \textcolor{red}{\textbf{\emph{0.7061}}} \\
\hline
     \textit{Contraction} & \textcolor{red}{\textbf{\emph{7.677}}} & 7.6502 & 7.6311 & 7.6246 & 7.6099 & 7.6022 & 7.5772 & 7.5675 & 7.5622 & 7.5541 & 7.5541 \\
\hline
    \# \textit{Inter-edges} & 64.4739 & 49.5072 & 45.0904 & 43.958 & 42.9969 & 42.2844 & 41.4681 & 40.9854 & 40.8404 & \textcolor{red}{\textbf{\emph{40.3328}}} & \textcolor{red}{\textbf{\emph{40.3328}}} \\
\hline
     \textit{Expansion} & 4.9436 & 3.8123 & 3.4761 & 3.3827 & 3.3049 & 3.2563 & 3.1943 & 3.1604 & 3.1516 & \textcolor{red}{\textbf{\emph{3.1237}}} & \textcolor{red}{\textbf{\emph{3.1237}}} \\
\hline
     \textit{Conductance} & 0.3838 & 0.3303 & 0.3126 & 0.3074 & 0.3032 & 0.3004 & 0.2973 & 0.2955 & 0.295 & \textcolor{red}{\textbf{\emph{0.2933}}} & \textcolor{red}{\textbf{\emph{0.2933}}} \\
\hline \hline
\end{tabular}
\end{table*}

\begin{table*}[!t]
\caption{The values of the metrics with the first version of belonging coefficient and the second version of belonging function on American college football network.}
\label{tab:football_1_2}
\vspace{-0.9em}
\centering
\begin{tabular}{c||c|c|c|c|c|c|c|c|c|c|c}
\hline \hline
     SLPA threshold $r$ & 0.01 & 0.05 & 0.1 & 0.15 & 0.2 & 0.25 & 0.3 & 0.35 & 0.4 & 0.45 & 0.5 \\
\hline
     $Q_{ov}$ & 0.5392 & 0.5657 & 0.5732 & 0.5746 & 0.5756 & 0.5762 & 0.5762 & 0.5765 & \textcolor{red}{\textbf{\emph{0.5766}}} & 0.5764 & 0.5764 \\
\hline
     $Q_{ov}^L$ & 0.66802 & 0.7019 & 0.71254 & 0.71519 & 0.71691 & 0.71811 & 0.71915 & 0.71933 & 0.71941 & \textcolor{red}{\textbf{\emph{0.71943}}} & \textcolor{red}{\textbf{\emph{0.71943}}} \\
\hline
     $Q_{ds}^{ov}$ & 0.3224 & 0.3588 & 0.3689 & 0.3714 & 0.3732 & 0.3742 & 0.374 & 0.3743 & 0.3742 & \textcolor{red}{\textbf{\emph{0.3747}}} & \textcolor{red}{\textbf{\emph{0.3747}}} \\
\hline
    \# \textit{Intra-edges} & 51.8519 & 53.4372 & 53.995 & 54.1026 & 54.1772 & 54.2286 & 54.2151 & 54.226 & 54.2304 & \textcolor{red}{\textbf{\emph{54.245}}} & \textcolor{red}{\textbf{\emph{54.245}}} \\
\hline
     \textit{Intra-density} & 0.6409 & 0.6869 & 0.6967 & 0.7004 & 0.7028 & 0.7037 & 0.7038 & 0.7048 & 0.7045 & \textcolor{red}{\textbf{\emph{0.7061}}} & \textcolor{red}{\textbf{\emph{0.7061}}} \\
\hline
     \textit{Contraction} & 7.0904 & 7.4155 & 7.5041 & 7.5262 & 7.5437 & 7.5513 & 7.5484 & 7.5503 & 7.5498 & \textcolor{red}{\textbf{\emph{7.5541}}} & \textcolor{red}{\textbf{\emph{7.5541}}} \\
\hline
    \# \textit{Inter-edges} & 46.4812 & 41.9484 & 40.8328 & 40.6176 & 40.4685 & 40.3656 & 40.3926 & 40.3707 & 40.3619 & \textcolor{red}{\textbf{\emph{40.3328}}} & \textcolor{red}{\textbf{\emph{40.3328}}} \\
\hline
     \textit{Expansion} & 3.5642 & 3.2499 & 3.1639 & 3.1445 & 3.1313 & 3.1246 & 3.1263 & 3.125 & 3.1252 & \textcolor{red}{\textbf{\emph{3.1237}}} & \textcolor{red}{\textbf{\emph{3.1237}}} \\
\hline
     \textit{Conductance} & 0.3358 & 0.3056 & 0.2975 & 0.2955 & 0.2941 & 0.2935 & 0.2936 & 0.2935 & 0.2935 & \textcolor{red}{\textbf{\emph{0.2933}}} & \textcolor{red}{\textbf{\emph{0.2933}}} \\
\hline \hline
\end{tabular}
\end{table*}

\begin{table*}[!t]
\caption{The values of the metrics with the second version of belonging coefficient and the first version of belonging function on American college football network.}
\label{tab:football_2_1}
\vspace{-0.9em}
\centering
\begin{tabular}{c||c|c|c|c|c|c|c|c|c|c|c}
\hline \hline
     SLPA threshold $r$ & 0.01 & 0.05 & 0.1 & 0.15 & 0.2 & 0.25 & 0.3 & 0.35 & 0.4 & 0.45 & 0.5 \\
\hline
     $Q_{ov}$ & 0.54283 & 0.56755 & 0.57376 & 0.57489 & 0.57545 & 0.57587 & 0.57611 & 0.57638 & \textcolor{red}{\textbf{\emph{0.57643}}} & 0.57636 & 0.57636 \\
\hline
     $Q_{ov}^L$ & 0.7002 & 0.7117 & 0.7168 & 0.7178 & 0.7188 & 0.7191 & \textcolor{red}{\textbf{\emph{0.7198}}} & 0.7197 & 0.7196 & 0.7194 & 0.7194 \\
\hline
     $Q_{ds}^{ov}$ & 0.3073 & 0.349 & 0.3627 & 0.3666 & 0.3694 & 0.3712 & 0.3724 & 0.3736 & 0.3737 & \textcolor{red}{\textbf{\emph{0.3747}}} & \textcolor{red}{\textbf{\emph{0.3747}}} \\
\hline
    \# \textit{Intra-edges} & \textcolor{red}{\textbf{\emph{56.4603}}} & 55.2278 & 54.9919 & 54.9048 & 54.7527 & 54.6671 & 54.4875 & 54.3979 & 54.3628 & 54.245 & 54.245 \\
\hline
     \textit{Intra-density} & 0.5979 & 0.6629 & 0.6818 & 0.6888 & 0.6947 & 0.6973 & 0.7004 & 0.7032 & 0.7032 & \textcolor{red}{\textbf{\emph{0.7061}}} & \textcolor{red}{\textbf{\emph{0.7061}}} \\
\hline
     \textit{Contraction} & \textcolor{red}{\textbf{\emph{7.762}}} & 7.672 & 7.644 & 7.6331 & 7.6157 & 7.6064 & 7.5787 & 7.5682 & 7.5629 & 7.5541 & 7.5541 \\
\hline
    \# \textit{Inter-edges} & 63.3502 & 49.2467 & 44.975 & 43.881 & 42.9432 & 42.256 & 41.4555 & 40.9783 & 40.836 & \textcolor{red}{\textbf{\emph{40.3328}}} & \textcolor{red}{\textbf{\emph{40.3328}}}  \\
\hline
     \textit{Expansion} & 4.8765 & 3.7943 & 3.4698 & 3.3786 & 3.3021 & 3.2553 & 3.1942 & 3.1603 & 3.1515 & \textcolor{red}{\textbf{\emph{3.1237}}} & \textcolor{red}{\textbf{\emph{3.1237}}}  \\
\hline
     \textit{Conductance} & 0.3781 & 0.3287 & 0.3119 & 0.3069 & 0.3029 & 0.3003 & 0.2973 & 0.2954 & 0.295 & \textcolor{red}{\textbf{\emph{0.2933}}} & \textcolor{red}{\textbf{\emph{0.2933}}} \\
\hline \hline
\end{tabular}
\end{table*}

\begin{table*}[!t]
\caption{The values of the metrics with the second version of belonging coefficient and the second version of belonging function on American college football network.}
\label{tab:football_2_2}
\vspace{-0.9em}
\centering
\begin{tabular}{c||c|c|c|c|c|c|c|c|c|c|c}
\hline \hline
     SLPA threshold $r$ & 0.01 & 0.05 & 0.1 & 0.15 & 0.2 & 0.25 & 0.3 & 0.35 & 0.4 & 0.45 & 0.5 \\
\hline
     $Q_{ov}$ & 0.5497 & 0.5682 & 0.5741 & 0.5752 & 0.5761 & 0.5763 & 0.5763 & 0.5765 & \textcolor{red}{\textbf{\emph{0.5766}}} & 0.5764 & 0.5764 \\
\hline
     $Q_{ov}^L$ & 0.7002 & 0.7117 & 0.7168 & 0.7178 & 0.7188 & 0.7191 & \textcolor{red}{\textbf{\emph{0.7198}}} & 0.7197 & 0.7196 & 0.7194 & 0.7194 \\
\hline
     $Q_{ds}^{ov}$ & 0.3359 & 0.3623 & 0.3706 & 0.3726 & 0.374 & 0.3746 & 0.3742 & 0.3744 & 0.3743 & \textcolor{red}{\textbf{\emph{0.3747}}} & \textcolor{red}{\textbf{\emph{0.3747}}} \\
\hline
    \# \textit{Intra-edges} & 52.8233 & 53.6506 & 54.0812 & 54.1625 & 54.218 & \textcolor{red}{\textbf{\emph{54.2494}}} & 54.2263 & 54.2332 & 54.2348 & 54.245 & 54.245 \\
\hline
     \textit{Intra-density} & 0.66 & 0.6912 & 0.6996 & 0.7023 & 0.7041 & 0.7047 & 0.7042 & 0.7049 & 0.7046 & \textcolor{red}{\textbf{\emph{0.7061}}} & \textcolor{red}{\textbf{\emph{0.7061}}} \\
\hline
     \textit{Contraction} & 7.2247 & 7.4483 & 7.5191 & 7.5358 & 7.5504 & \textcolor{red}{\textbf{\emph{7.5553}}} & 7.5499 & 7.5512 & 7.5505 & 7.5541 & 7.5541 \\
\hline
    \# \textit{Inter-edges} & 44.5385 & 41.5216 & 40.6605 & 40.4978 & 40.3868 & \textcolor{red}{\textbf{\emph{40.324}}} & 40.3703 & 40.3564 & 40.3532 & 40.3328 & 40.3328 \\
\hline
     \textit{Expansion} & 3.4334 & 3.2197 & 3.1522 & 3.1372 & 3.1262 & \textcolor{red}{\textbf{\emph{3.1221}}} & 3.1254 & 3.1244 & 3.1247 & 3.1237 & 3.1237 \\
\hline
     \textit{Conductance} & 0.32342 & 0.3027 & 0.2963 & 0.2948 & 0.2936 & \textcolor{red}{\textbf{\emph{0.2932}}} & 0.2935 & 0.2935 & 0.2935 & 0.2933 & 0.2933 \\
\hline \hline
\end{tabular}
\vspace{-0.2em}
\end{table*}

Tables~\ref{tab:karate_1_1}-\ref{tab:karate_2_2}, Tables~\ref{tab:football_1_1}-\ref{tab:football_2_2}, Tables~\ref{tab:jazz_1_1}-\ref{tab:jazz_2_2}, and Tables~\ref{tab:pgp_1_1}-\ref{tab:pgp_2_2} present the values of $Q_{ov}$, $Q_{ov}^L$, $Q_{ds}^{ov}$, and the six metrics for the community detection results obtained by running SLPA with threshold $r$ varying from 0.01 to 0.5 on Zachary's karate club network, American college football network, Jazz musicians network, and PGP network, respectively. For each network, there are four tables corresponding to two versions of the belonging coefficient and two versions of the belonging function. In each of these tables, red italic font denotes the best value of each metric among the cases with different values of threshold $r$. Notice that for each network, the values of $Q_{ov}^L$ in the first table are the same as those in the second table and the values of $Q_{ov}^L$ in the third table are equal to those in the fourth table. This is because $Q_{ov}^L$ has its own belonging function shown in Equation~(\ref{eq:Qov_L_bf}).

Tables~\ref{tab:karate_1_1}-\ref{tab:karate_2_2} show that the nine metrics with the second version of the belonging function are better since the number of consistent metrics corresponding to the best value of threshold $r$ ($r=0.45$) for SLPA is at $7$ the largest. Tables~\ref{tab:football_1_1}-\ref{tab:football_2_2} indicate that the nine metrics with the first version of the belonging coefficient and the second version of the belonging function are the best among the four combinations since there are eight metrics, except $Q_{ov}$, consistent with each other. Tables~\ref{tab:jazz_1_1}-\ref{tab:jazz_2_2} suggest that the nine metrics with the first version of the belonging coefficient are better. Tables~\ref{tab:pgp_1_1}-\ref{tab:pgp_2_2} demonstrate that the nine metrics with the first version of the belonging coefficient and the second version of the belonging function are the best. In conclusion, the nine metrics with the first version of the belonging coefficient and the second version of the belonging function are the best among the four combinations. This is because in this case the number of consistent metrics corresponding to the best value of threshold $r$ for SLPA is the largest for all four networks. Thus, we conclude that the first version the belonging coefficient is better. It means that for converting crisp overlapping to fuzzy overlapping the belonging coefficient of a node to a community should be the reciprocal of the number of communities in which this node participates. When the relationship between a node and the communities to which it belongs is binary, there is no information about the strength of the membership. In this case, it is intuitive and reasonable to assign a node to its communities using equal belonging coefficients. We also determined that the second version of the belonging function is better. It means that the probability of the event that two nodes belong to the same community should be the product, not the average, of their belonging coefficients to that community. In addition, $Q_{ov}=Q_{ov}'$ when $f(a_{i,c}, a_{j,c})=a_{i,c}a_{j,c}$ proved in Subsection~\ref{sec:ov_Q}, which is another way of showing that the second version of the belonging function is much more suitable for use in the metric than the first.

\begin{table*}[!t]
\caption{The values of the metrics with the first version of belonging coefficient and the first version of belonging function on Jazz Musicians Network.}
\label{tab:jazz_1_1}
\vspace{-0.9em}
\centering
\begin{tabular}{c||c|c|c|c|c|c|c|c|c|c|c}
\hline \hline
     SLPA threshold $r$ & 0.01 & 0.05 & 0.1 & 0.15 & 0.2 & 0.25 & 0.3 & 0.35 & 0.4 & 0.45 & 0.5 \\
\hline
     $Q_{ov}$ & 0.3339 & 0.3535 & 0.3598 & 0.3622 & 0.3637 & 0.3655 & 0.366 & 0.367 & 0.3679 & 0.3686 & \textcolor{red}{\textbf{\emph{0.3696}}} \\
\hline
     $Q_{ov}^L$ & 0.689 & 0.69623 & 0.69812 & 0.6984 & 0.69888 & 0.69944 & 0.70008 & 0.70054 & 0.70109 & 0.70108 & \textcolor{red}{\textbf{\emph{0.70114}}} \\
\hline
     $Q_{ds}^{ov}$ & 0.1896 & 0.1986 & 0.2006 & 0.2018 & 0.2025 & 0.2033 & 0.2039 & 0.2044 & 0.2049 & 0.2049 & \textcolor{red}{\textbf{\emph{0.2052}}} \\
\hline
    \# \textit{Intra-edges} & \textcolor{red}{\textbf{\emph{858.5208}}} & 745.7104 & 742.0708 & 696.7875 & 695.8667 & 694.65 & 694.3917 & 693.8208 & 693.2646 & 692.2438 & 691.225 \\
\hline
     \textit{Intra-density} & 0.353 & 0.4367 & 0.4655 & 0.4915 & 0.4923 & 0.4933 & 0.4941 & 0.4947 & 0.4952 & 0.4953 & \textcolor{red}{\textbf{\emph{0.4957}}} \\
\hline
     \textit{Contraction} & \textcolor{red}{\textbf{\emph{20.5048}}} & 18.3409 & 18.1895 & 17.4338 & 17.416 & 17.3818 & 17.3751 & 17.3661 & 17.3546 & 17.3242 & 17.2971 \\
\hline
    \# \textit{Inter-edges} & 321.4583 & 257.8458 & 243.1333 & 232.2917 & 229.5 & 225.7083 & 224.5917 & 223.1083 & 221.5875 & 219.2208 & \textcolor{red}{\textbf{\emph{217.05}}} \\
\hline
     \textit{Expansion} & 5.0762 & 4.2512 & 3.9404 & 3.7894 & 3.7144 & 3.6586 & 3.6429 & 3.5927 & 3.5482 & 3.4907 & \textcolor{red}{\textbf{\emph{3.4577}}} \\
\hline
     \textit{Conductance} & 0.2214 & 0.2327 & 0.2201 & 0.2213 & 0.2169 & 0.2156 & 0.2151 & 0.2113 & 0.2071 & 0.2027 & \textcolor{red}{\textbf{\emph{0.202}}} \\
\hline \hline
\end{tabular}
\vspace{-0.2em}
\end{table*}

\begin{table*}[!t]
\caption{The values of the metrics with the first version of belonging coefficient and the second version of belonging function on Jazz Musicians Network.}
\label{tab:jazz_1_2}
\vspace{-0.9em}
\centering
\begin{tabular}{c||c|c|c|c|c|c|c|c|c|c|c}
\hline \hline
     SLPA threshold $r$ & 0.01 & 0.05 & 0.1 & 0.15 & 0.2 & 0.25 & 0.3 & 0.35 & 0.4 & 0.45 & 0.5 \\
\hline
     $Q_{ov}$ & 0.3591 & 0.36561 & 0.36719 & 0.36743 & 0.36807 & 0.36848 & 0.36875 & 0.36915 & 0.36956 & 0.36948 & \textcolor{red}{\textbf{\emph{0.36962}}} \\
\hline
     $Q_{ov}^L$ & 0.689 & 0.69623 & 0.69812 & 0.6984 & 0.69888 & 0.69944 & 0.70008 & 0.70054 & 0.70109 & 0.70108 & \textcolor{red}{\textbf{\emph{0.70114}}} \\
\hline
     $Q_{ds}^{ov}$ & 0.19495 & 0.20144 & 0.20253 & 0.20357 & 0.20401 & 0.20432 & 0.20446 & 0.20471 & 0.20497 & 0.20507 & \textcolor{red}{\textbf{\emph{0.20516}}} \\
\hline
    \# \textit{Intra-edges} & \textcolor{red}{\textbf{\emph{830.0938}}} & 731.7604 & 733.4229 & 690.4646 & 690.6271 & 690.9042 & 691.0458 & 691.0875 & 691.1833 & 691.2083 & 691.225 \\
\hline
     \textit{Intra-density} & 0.36491 & 0.4435 & 0.46926 & 0.49481 & 0.49514 & 0.49525 & 0.49525 & 0.49543 & 0.49542 & 0.49555 & \textcolor{red}{\textbf{\emph{0.49569}}} \\
\hline
     \textit{Contraction} & \textcolor{red}{\textbf{\emph{19.5917}}} & 17.8407 & 17.8602 & 17.1896 & 17.2187 & 17.229 & 17.2342 & 17.2556 & 17.2767 & 17.2929 & 17.2971 \\
\hline
    \# \textit{Inter-edges} & 259.2125 & 227.3792 & 224.0542 & 218.5708 & 218.2458 & 217.6917 & 217.4083 & 217.325 & 217.1333 & 217.0833 & \textcolor{red}{\textbf{\emph{217.05}}} \\
\hline
     \textit{Expansion} & 4.0312 & 3.6738 & 3.5718 & 3.5203 & 3.501 & 3.494 & 3.4905 & 3.4777 & 3.4659 & 3.4585 & \textcolor{red}{\textbf{\emph{3.4577}}} \\
\hline
     \textit{Conductance} & 0.20298 & 0.22321 & 0.21467 & 0.21752 & 0.21321 & 0.21285 & 0.21268 & 0.20862 & 0.20529 & 0.20209 & \textcolor{red}{\textbf{\emph{0.20202}}} \\
\hline \hline
\end{tabular}
\vspace{-0.2em}
\end{table*}

\begin{table*}[!t]
\caption{The values of the metrics with the second version of belonging coefficient and the first version of belonging function on Jazz Musicians Network.}
\label{tab:jazz_2_1}
\vspace{-0.9em}
\centering
\begin{tabular}{c||c|c|c|c|c|c|c|c|c|c|c}
\hline \hline
     SLPA threshold $r$ & 0.01 & 0.05 & 0.1 & 0.15 & 0.2 & 0.25 & 0.3 & 0.35 & 0.4 & 0.45 & 0.5 \\
\hline
     $Q_{ov}$ & 0.3349 & 0.3537 & 0.3599 & 0.3622 & 0.3637 & 0.3656 & 0.366 & 0.367 & 0.3679 & 0.3686 & \textcolor{red}{\textbf{\emph{0.3696}}} \\
\hline
     $Q_{ov}^L$ & 0.6989 & 0.6998 & 0.7004 & 0.7007 & 0.701 & 0.701 & 0.7009 & 0.7011 & \textcolor{red}{\textbf{\emph{0.7013}}} & 0.7012 & 0.7011 \\
\hline
     $Q_{ds}^{ov}$ & 0.1903 & 0.1987 & 0.2006 & 0.2018 & 0.2025 & 0.2033 & 0.2039 & 0.2044 & 0.2049 & 0.2049 & \textcolor{red}{\textbf{\emph{0.2052}}} \\
\hline
    \# \textit{Intra-edges} & \textcolor{red}{\textbf{\emph{859.6312}}} & 745.9458 & 742.1291 & 696.8309 & 695.9008 & 694.6762 & 694.4036 & 693.8312 & 693.2664 & 692.2457 & 691.225 \\
\hline
     \textit{Intra-density} & 0.3559 & 0.4402 & 0.4655 & 0.4914 & 0.4923 & 0.4932 & 0.4941 & 0.4947 & 0.4952 & 0.4953 & \textcolor{red}{\textbf{\emph{0.4957}}} \\
\hline
     \textit{Contraction} & \textcolor{red}{\textbf{\emph{20.5333}}} & 18.3469 & 18.1877 & 17.4327 & 17.4158 & 17.3819 & 17.3741 & 17.3659 & 17.3547 & 17.3243 & 17.2971 \\
\hline
    \# \textit{Inter-edges} & 319.2376 & 257.3751 & 243.0169 & 232.2048 & 229.4316 & 225.6557 & 224.5677 & 223.0876 & 221.5837 & 219.2169 & \textcolor{red}{\textbf{\emph{217.05}}} \\
\hline
     \textit{Expansion} & 5.0709 & 4.25 & 3.9472 & 3.7945 & 3.7167 & 3.6612 & 3.6451 & 3.5917 & 3.5479 & 3.4905 & \textcolor{red}{\textbf{\emph{3.4577}}} \\
\hline
     \textit{Conductance} & 0.2207 & 0.2318 & 0.2209 & 0.2219 & 0.2172 & 0.2159 & 0.2154 & 0.2113 & 0.207 & 0.2026 & \textcolor{red}{\textbf{\emph{0.202}}} \\
\hline \hline
\end{tabular}
\vspace{-0.2em}
\end{table*}

\begin{table*}[!t]
\caption{The values of the metrics with the second version of belonging coefficient and the second version of belonging function on Jazz Musicians Network.}
\label{tab:jazz_2_2}
\vspace{-0.9em}
\centering
\begin{tabular}{c||c|c|c|c|c|c|c|c|c|c|c}
\hline \hline
     SLPA threshold $r$ & 0.01 & 0.05 & 0.1 & 0.15 & 0.2 & 0.25 & 0.3 & 0.35 & 0.4 & 0.45 & 0.5 \\
\hline
     $Q_{ov}$ & 0.36115 & 0.36605 & 0.36736 & 0.36758 & 0.36822 & 0.36861 & 0.36881 & 0.36922 & 0.36958 & 0.3695 & \textcolor{red}{\textbf{\emph{0.36962}}} \\
\hline
     $Q_{ov}^L$ & 0.6989 & 0.6998 & 0.7004 & 0.7007 & 0.701 & 0.701 & 0.7009 & 0.7011 & \textcolor{red}{\textbf{\emph{0.7013}}} & 0.7012 & 0.7011 \\
\hline
     $Q_{ds}^{ov}$ & 0.1964 & 0.2016 & 0.2026 & 0.2036 & 0.2041 & 0.2044 & 0.2045 & 0.2047 & 0.205 & 0.2051 & \textcolor{red}{\textbf{\emph{0.2052}}} \\
\hline
    \# \textit{Intra-edges} & \textcolor{red}{\textbf{\emph{832.3217}}} & 732.2389 & 733.5414 & 690.5535 & 690.6971 & 690.9578 & 691.0702 & 691.1083 & 691.1871 & 691.2123 & 691.225 \\
\hline
     \textit{Intra-density} & 0.3741 & 0.4562 & 0.4691 & 0.4947 & 0.4951 & 0.4952 & 0.4952 & 0.4954 & 0.4954 & 0.4956 & \textcolor{red}{\textbf{\emph{0.4957}}} \\
\hline
     \textit{Contraction} & \textcolor{red}{\textbf{\emph{19.6306}}} & 17.8419 & 17.8522 & 17.1842 & 17.2169 & 17.2274 & 17.2309 & 17.2562 & 17.2771 & 17.2933 & 17.2971 \\
\hline
    \# \textit{Inter-edges} & 254.7564 & 226.4221 & 223.8172 & 218.3929 & 218.1058 & 217.5843 & 217.3596 & 217.2834 & 217.1258 & 217.0755 & \textcolor{red}{\textbf{\emph{217.05}}} \\
\hline
     \textit{Expansion} & 3.9449 & 3.6284 & 3.5701 & 3.5192 & 3.4999 & 3.4933 & 3.49 & 3.4765 & 3.4657 & 3.4583 & \textcolor{red}{\textbf{\emph{3.4577}}} \\
\hline
     \textit{Conductance} & 0.1985 & 0.2187 & 0.2167 & 0.219 & 0.2139 & 0.2135 & 0.2134 & 0.2086 & 0.2053 & 0.2021 & \textcolor{red}{\textbf{\emph{0.202}}} \\
\hline \hline
\end{tabular}
\vspace{-0.8em}
\end{table*}

Moreover, for the other three combinations of the belonging coefficient and the belonging function, $Q_{ds}^{ov}$ is always consistent with major metrics while sometimes $Q_{ov}$ and sometimes $Q_{ov}^L$ are not. It follows that $Q_{ds}^{ov}$ has an advantage over both $Q_{ov}$ and $Q_{ov}^L$. Thus, we conclude that for the networks that were used in our experiments $Q_{ds}^{ov}$ with the first version of the belonging coefficient and the second version of the belonging function surpassed other metrics quantifying the quality of overlapping community structures. Furthermore, among all overlapping extensions of modularity, we recommend using $Q_{ov}$ with the first version of the belonging coefficient and the second version of the belonging function for its effectiveness and simplicity.

\section{Conclusion}
\label{sec:conclusion}
In this paper, we determined which version of the belonging coefficient and which version of the belonging function are better for measuring quality of overlapping communities. We also showed which extension of modularity performed best on our tests among all overlapping extensions of modularity. Moreover, we proposed an overlapping extension for modularity density based on the overlapping extensions of modularity. We concluded that overlapping modularity density is the best for measuring quality of overlapping community structure.

\section*{Acknowledgment}
This work was supported in part by the Army Research Laboratory under Cooperative Agreement Number W911NF-09-2-0053 and by the the Office of Naval Research Grant No. N00014-09-1-0607. The views and conclusions contained in this paper are those of the authors and should not be interpreted as representing the official policies either expressed or implied of the Army Research Laboratory or the U.S. Government.

\begin{table*}[!t]
\caption{The values of the metrics with the first version of belonging coefficient and the first version of belonging function on PGP Network.}
\label{tab:pgp_1_1}
\vspace{-0.9em}
\centering
\begin{tabular}{c||c|c|c|c|c|c|c|c|c|c|c}
\hline \hline
     SLPA threshold $r$ & 0.01 & 0.05 & 0.1 & 0.15 & 0.2 & 0.25 & 0.3 & 0.35 & 0.4 & 0.45 & 0.5 \\
\hline
     $Q_{ov}$ & 0.8121 & \textcolor{red}{\textbf{\emph{0.8143}}} & 0.8121 & 0.8096 & 0.8077 & 0.8057 & 0.8038 & 0.8014 & 0.7997 & 0.7982 & 0.7972 \\
\hline
     $Q_{ov}^L$ & 0.7456 & 0.7849 & 0.7969 & 0.8024 & 0.8065 & 0.8087 & 0.8105 & 0.8112 & 0.8124 & 0.8135 & \textcolor{red}{\textbf{\emph{0.8137}}} \\
\hline
     $Q_{ds}^{ov}$ & 0.1312 & 0.1422 & 0.1474 & 0.1506 & 0.1532 & 0.1554 & 0.1573 & 0.1585 & 0.1595 & 0.1602 & \textcolor{red}{\textbf{\emph{0.1603}}} \\
\hline
    \# \textit{Intra-edges} & \textcolor{red}{\textbf{\emph{24.4988}}} & 24.0138 & 23.508 & 22.9699 & 22.5532 & 22.1654 & 21.8903 & 21.5908 & 21.4259 & 21.3164 & 21.272 \\
\hline
     \textit{Intra-density} & 0.3571 & 0.3737 & 0.3822 & 0.3894 & 0.3954 & 0.4011 & 0.4049 & 0.4086 & 0.4113 & \textcolor{red}{\textbf{\emph{0.4129}}} & 0.4127 \\
\hline
     \textit{Contraction} & \textcolor{red}{\textbf{\emph{2.51}}} & 2.4633 & 2.44 & 2.4174 & 2.4 & 2.3835 & 2.3698 & 2.3555 & 2.3453 & 2.337 & 2.3331 \\
\hline
    \# \textit{Inter-edges} & 22.3009 & 16.2737 & 14.1732 & 12.9528 & 12.0829 & 11.4285 & 10.9102 & 10.4893 & 10.1373 & 9.8355 & \textcolor{red}{\textbf{\emph{9.7283}}} \\
\hline
     \textit{Expansion} & 1.6656 & 1.0983 & 0.9085 & 0.8346 & 0.7731 & 0.7283 & 0.6806 & 0.6499 & 0.6186 & 0.591 & \textcolor{red}{\textbf{\emph{0.5797}}} \\
\hline
     \textit{Conductance} & 0.297 & 0.2535 & 0.2342 & 0.2259 & 0.2189 & 0.2138 & 0.2079 & 0.204 & 0.1992 & 0.1948 & \textcolor{red}{\textbf{\emph{0.1929}}} \\
\hline \hline
\end{tabular}
\vspace{-0.1em}
\end{table*}

\begin{table*}[!t]
\caption{The values of the metrics with the first version of belonging coefficient and the second version of belonging function on PGP Network.}
\label{tab:pgp_1_2}
\vspace{-0.9em}
\centering
\begin{tabular}{c||c|c|c|c|c|c|c|c|c|c|c}
\hline \hline
     SLPA threshold $r$ & 0.01 & 0.05 & 0.1 & 0.15 & 0.2 & 0.25 & 0.3 & 0.35 & 0.4 & 0.45 & 0.5 \\
\hline
     $Q_{ov}$ & 0.7447 & 0.7744 & 0.7832 & 0.7873 & 0.7904 & 0.7924 & 0.7941 & 0.7948 & 0.796 & 0.797 & \textcolor{red}{\textbf{\emph{0.7972}}} \\
\hline
     $Q_{ov}^L$ & 0.7456 & 0.7849 & 0.7969 & 0.8024 & 0.8065 & 0.8087 & 0.8105 & 0.8112 & 0.8124 & 0.8135 & \textcolor{red}{\textbf{\emph{0.8137}}} \\
\hline
     $Q_{ds}^{ov}$ & 0.1298 & 0.1409 & 0.146 & 0.1493 & 0.1521 & 0.1543 & 0.1564 & 0.1578 & 0.1591 & 0.16 & \textcolor{red}{\textbf{\emph{0.1603}}} \\
\hline
    \# \textit{Intra-edges} & 22.3576 & \textcolor{red}{\textbf{\emph{22.7808}}} & 22.6356 & 22.3116 & 22.0523 & 21.7862 & 21.6151 & 21.4051 & 21.3214 & 21.2838 & 21.272 \\
\hline
     \textit{Intra-density} & 0.3648 & 0.3796 & 0.3867 & 0.393 & 0.3983 & 0.4035 & 0.4066 & 0.4099 & 0.4121 & \textcolor{red}{\textbf{\emph{0.4132}}} & 0.4127 \\
\hline
     \textit{Contraction} & 2.2332 & 2.297 & 2.321 & 2.3226 & 2.3248 & 2.3239 & 2.3272 & 2.3254 & 2.3282 & 2.3315 & \textcolor{red}{\textbf{\emph{2.3331}}} \\
\hline
    \# \textit{Inter-edges} & 14.051 & 12.0356 & 11.325 & 10.8721 & 10.5311 & 10.2765 & 10.0899 & 9.9485 & 9.8337 & 9.7465 & \textcolor{red}{\textbf{\emph{9.7283}}} \\
\hline
     \textit{Expansion} & 0.9636 & 0.7528 & 0.6839 & 0.6611 & 0.6403 & 0.6265 & 0.6105 & 0.6026 & 0.5922 & 0.583 & \textcolor{red}{\textbf{\emph{0.5797}}} \\
\hline
     \textit{Conductance} & 0.2632 & 0.2291 & 0.2155 & 0.211 & 0.2068 & 0.2042 & 0.2007 & 0.1991 & 0.1966 & 0.194 & \textcolor{red}{\textbf{\emph{0.1929}}} \\
\hline \hline
\end{tabular}
\vspace{-0.1em}
\end{table*}

\begin{table*}[!t]
\caption{The values of the metrics with the second version of belonging coefficient and the first version of belonging function on PGP Network.}
\label{tab:pgp_2_1}
\vspace{-0.9em}
\centering
\begin{tabular}{c||c|c|c|c|c|c|c|c|c|c|c}
\hline \hline
     SLPA threshold $r$ & 0.01 & 0.05 & 0.1 & 0.15 & 0.2 & 0.25 & 0.3 & 0.35 & 0.4 & 0.45 & 0.5 \\
\hline
     $Q_{ov}$ & \textcolor{red}{\textbf{\emph{0.8212}}} & 0.8187 & 0.8151 & 0.8119 & 0.8095 & 0.807 & 0.8047 & 0.802 & 0.8 & 0.7983 & 0.7972 \\
\hline
     $Q_{ov}^L$ & 0.7952 & 0.8118 & 0.8158 & 0.8167 & \textcolor{red}{\textbf{\emph{0.8172}}} & 0.8167 & 0.8162 & 0.815 & 0.8144 & 0.8141 & 0.8137 \\
\hline
     $Q_{ds}^{ov}$ & 0.1324 & 0.1428 & 0.1477 & 0.1508 & 0.1535 & 0.1555 & 0.1574 & 0.1586 & 0.1595 & 0.1602 & \textcolor{red}{\textbf{\emph{0.1603}}} \\
\hline
    \# \textit{Intra-edges} & \textcolor{red}{\textbf{\emph{24.781}}} & 24.1466 & 23.5983 & 23.0377 & 22.604 & 22.2024 & 21.9162 & 21.6081 & 21.4355 & 21.3193 & 21.272 \\
\hline
     \textit{Intra-density} & 0.3592 & 0.3754 & 0.3834 & 0.3905 & 0.3962 & 0.4018 & 0.4053 & 0.409 & 0.4116 & \textcolor{red}{\textbf{\emph{0.413}}} & 0.4127 \\
\hline
     \textit{Contraction} & \textcolor{red}{\textbf{\emph{2.5264}}} & 2.4728 & 2.4465 & 2.4226 & 2.4041 & 2.3863 & 2.3719 & 2.357 & 2.3462 & 2.3373 & 2.3331 \\
\hline
    \# \textit{Inter-edges} & 21.7366 & 16.0079 & 13.9926 & 12.8173 & 11.9813 & 11.3545 & 10.8583 & 10.4546 & 10.1182 & 9.8297 & \textcolor{red}{\textbf{\emph{9.7283}}} \\
\hline
     \textit{Expansion} & 1.7537 & 1.1229 & 0.9162 & 0.844 & 0.7797 & 0.7326 & 0.6817 & 0.6495 & 0.6182 & 0.5907 & \textcolor{red}{\textbf{\emph{0.5797}}} \\
\hline
     \textit{Conductance} & 0.2943 & 0.2516 & 0.2327 & 0.2247 & 0.218 & 0.213 & 0.2073 & 0.2036 & 0.199 & 0.1947 & \textcolor{red}{\textbf{\emph{0.1929}}} \\
\hline \hline
\end{tabular}
\vspace{-0.1em}
\end{table*}

\begin{table*}[!t]
\caption{The values of the metrics with the second version of belonging coefficient and the second version of belonging function on PGP Network.}
\label{tab:pgp_2_2}
\vspace{-0.9em}
\centering
\begin{tabular}{c||c|c|c|c|c|c|c|c|c|c|c}
\hline \hline
     SLPA threshold $r$ & 0.01 & 0.05 & 0.1 & 0.15 & 0.2 & 0.25 & 0.3 & 0.35 & 0.4 & 0.45 & 0.5 \\
\hline
     $Q_{ov}$ & 0.761641 & 0.782874 & 0.789139 & 0.79193 & 0.793999 & 0.795024 & 0.795936 & 0.796058 & 0.796669 & \textcolor{red}{\textbf{\emph{0.797234}}} & 0.797232 \\
\hline
     $Q_{ov}^L$ & 0.7952 & 0.8118 & 0.8158 & 0.8167 & \textcolor{red}{\textbf{\emph{0.8172}}} & 0.8167 & 0.8162 & 0.815 & 0.8144 & 0.8141 & 0.8137 \\
\hline
     $Q_{ds}^{ov}$ & 0.1326 & 0.1423 & 0.147 & 0.15 & 0.1526 & 0.1547 & 0.1567 & 0.158 & 0.1592 & 0.1601 & \textcolor{red}{\textbf{\emph{0.1603}}} \\
\hline
    \# \textit{Intra-edges} & 22.8729 & \textcolor{red}{\textbf{\emph{23.0352}}} & 22.8104 & 22.4451 & 22.1531 & 21.8601 & 21.6673 & 21.4401 & 21.3407 & 21.2897 & 21.272 \\
\hline
     \textit{Intra-density} & 0.372 & 0.385 & 0.3909 & 0.3967 & 0.4014 & 0.4059 & 0.4083 & 0.4112 & 0.413 & \textcolor{red}{\textbf{\emph{0.4135}}} & 0.4127 \\
\hline
     \textit{Contraction} & 2.2454 & 2.3043 & 2.3259 & 2.3261 & 2.3272 & 2.3251 & 2.3279 & 2.3261 & 2.3287 & 2.3316 & \textcolor{red}{\textbf{\emph{2.3331}}} \\
\hline
    \# \textit{Inter-edges} & 13.0203 & 11.5268 & 10.9755 & 10.6052 & 10.3294 & 10.1286 & 9.9854 & 9.8784 & 9.7952 & 9.7347 & \textcolor{red}{\textbf{\emph{9.7283}}} \\
\hline
     \textit{Expansion} & 0.8828 & 0.7109 & 0.6539 & 0.6377 & 0.6214 & 0.6116 & 0.5998 & 0.595 & 0.5877 & 0.5814 & \textcolor{red}{\textbf{\emph{0.5797}}} \\
\hline
     \textit{Conductance} & 0.2528 & 0.2222 & 0.2102 & 0.2067 & 0.2033 & 0.2015 & 0.1987 & 0.1977 & 0.1957 & 0.1936 & \textcolor{red}{\textbf{\emph{0.1927}}} \\
\hline \hline
\end{tabular}
\vspace{-0.2em}
\end{table*}


\begin{thebibliography}{10}
\providecommand{\url}[1]{#1}
\csname url@samestyle\endcsname
\providecommand{\newblock}{\relax}
\providecommand{\bibinfo}[2]{#2}
\providecommand{\BIBentrySTDinterwordspacing}{\spaceskip=0pt\relax}
\providecommand{\BIBentryALTinterwordstretchfactor}{4}
\providecommand{\BIBentryALTinterwordspacing}{\spaceskip=\fontdimen2\font plus
\BIBentryALTinterwordstretchfactor\fontdimen3\font minus
  \fontdimen4\font\relax}
\providecommand{\BIBforeignlanguage}[2]{{%
\expandafter\ifx\csname l@#1\endcsname\relax
\typeout{** WARNING: IEEEtran.bst: No hyphenation pattern has been}%
\typeout{** loaded for the language `#1'. Using the pattern for}%
\typeout{** the default language instead.}%
\else
\language=\csname l@#1\endcsname
\fi
#2}}
\providecommand{\BIBdecl}{\relax}
\BIBdecl

\bibitem{UWDModularity}
M.~E.~J. Newman and M.~Girvan, ``Finding and evaluating community structure in
  networks,'' \emph{Phys. Rev. E}, vol.~69, p. 026113, Feb 2004.

\bibitem{CommunityReport}
S.~Fortunato, ``Community detection in graphs,'' \emph{Physics Reports}, vol.
  486, pp. 75--174, 2010.

\bibitem{NewmanGreedy}
M.~E.~J. Newman, ``Fast algorithm for detecting community structure in
  networks,'' \emph{Phys. Rev. E}, vol.~69, p. 066133, Jun 2004.

\bibitem{PNASModularity}
M.~E.~J. Newman, ``Modularity and community structure in networks,''
  \emph{Proceedings of the National Academy of Sciences}, vol. 103, no.~23, pp.
  8577--8582, 2006.

\bibitem{Overlapping_survey}
J.~Xie, S.~Kelley, and B.~K. Szymanski, ``Overlapping community detection in
  networks: The state-of-the-art and comparative study,'' \emph{ACM Comput.
  Surv.}, vol.~45, no.~4, pp. 43:1--43:35, Aug. 2013.

\bibitem{fuzzy_cmeans_Qov}
S.~Zhang, R.-S. Wang, and X.-S. Zhang, ``Identification of overlapping
  community structure in complex networks using fuzzy c-means clustering,''
  \emph{Physica A: Statistical Mechanics and its Applications}, vol. 374,
  no.~1, pp. 483 -- 490, 2007.

\bibitem{fuzzy_opt_Qov}
T.~Nepusz, A.~Petr\'oczi, L.~N\'egyessy, and F.~Bazs\'o, ``Fuzzy communities
  and the concept of bridgeness in complex networks,'' \emph{Phys. Rev. E},
  vol.~77, p. 016107, Jan 2008.

\bibitem{Qov_num_coms}
H.~Shen, X.~Cheng, K.~Cai, and M.-B. Hu, ``Detect overlapping and hierarchical
  community structure in networks,'' \emph{Physica A: Statistical Mechanics and
  its Applications}, vol. 388, no.~8, pp. 1706 -- 1712, 2009.

\bibitem{Qov_max_clique}
H.-W. Shen, X.-Q. Cheng, and J.-F. Guo, ``Quantifying and identifying the
  overlapping community structure in networks,'' \emph{Journal of Statistical
  Mechanics: Theory and Experiment}, vol. 2009, no.~07, p. P07042, 2009.

\bibitem{Qov_node_strength}
D.~Chen, M.~Shang, Z.~Lv, and Y.~Fu, ``Detecting overlapping communities of
  weighted networks via a local algorithm,'' \emph{Physica A: Statistical
  Mechanics and its Applications}, vol. 389, no.~19, pp. 4177 -- 4187, 2010.

\bibitem{Qov_edge_degree}
V.~Nicosia, G.~Mangioni, V.~Carchiolo, and M.~Malgeri, ``Extending the
  definition of modularity to directed graphs with overlapping communities,''
  \emph{Journal of Statistical Mechanics: Theory and Experiment}, vol. 2009,
  no.~03, p. P03024, 2009.

\bibitem{OverlappingQ_survey}
S.~Gregory, ``Fuzzy overlapping communities in networks,'' \emph{Journal of
  Statistical Mechanics: Theory and Experiment}, vol. 2011, no.~02, p. P02017,
  2011.

\bibitem{QdsConference}
M.~Chen, T.~Nguyen, and B.~K. Szymanski, ``On measuring the quality of a
  network community structure,'' in \emph{Proceedings of ASE/IEEE International
  Conference on Social Computing}, Washington, DC, USA, September 2013.

\bibitem{QdsJournal}
M.~Chen, T.~Nguyen, and B.~K. Szymanski, ``A new metric for quality of network
  community structure,'' \emph{ASE Human Journal}, vol.~2, no.~4, pp. 226--240,
  2013.

\bibitem{ov_lpa}
S.~Gregory, ``Finding overlapping communities in networks by label
  propagation,'' \emph{New Journal of Physics}, vol.~12, no.~10, p. 103018,
  2010.

\bibitem{ResolutionLimit}
S.~Fortunato and M.~Barth\`{e}lemy, ``Resolution limit in community
  detection,'' \emph{Proceedings of the National Academy of Sciences}, vol.
  104, no.~1, pp. 36--41, 2007.

\bibitem{SLPA2012}
J.~Xie and B.~K. Szymanski, ``Towards linear time overlapping community
  detection in social networks,'' in \emph{The 16th Pacific-Asia Conference on
  Knowledge Discovery and Data Mining (PAKDD)}, 2012, pp. 25--36.

\bibitem{karate}
W.~Zachary, ``An information flow model for conflict and fission in small
  groups,'' \emph{Journal of Anthropological Research}, vol.~33, pp. 452--473,
  1977.

\bibitem{football}
M.~Girvan and M.~E.~J. Newman, ``Community structure in social and biological
  networks,'' \emph{Proceedings of the National Academy of Sciences}, vol.~99,
  no.~12, pp. 7821--7826, 2002.

\bibitem{Jazz}
P.~Gleiser and L.~Danon, ``Community structure in jazz,'' \emph{Advances in
  Complex Systems}, vol.~06, no.~04, pp. 565--573, 2003.

\bibitem{PGPNetwork}
M.~Bogu\~{n}\'{a}, R.~Pastor-Satorras, A.~D\'{i}az-Guilera, and A.~Arenas,
  ``Models of social networks based on social distance attachment,''
  \emph{Phys. Rev. E}, vol.~70, p. 056122, Nov 2004.

\end{thebibliography}


\end{document}